\documentclass{article}
\usepackage{amsmath}
\usepackage{amsfonts}
\usepackage{amssymb}
\usepackage{graphicx}

\input{tcilatex}

\begin{document}

\title{Global analysis of the data from solar neutrinos having transition
magnetic moments together with KamLAND data}
\author{D. Yilmaz and A.U. Yilmazer \\
Department of Physics Engineering,Ankara University\\
Tandogan,06100, Ankara TURKEY}
\maketitle

\begin{abstract}
A global analysis of the solar neutrino data from all solar neutrino
experiments combined with the KamLAND data is presented assuming that the
solar neutrino deficit is due to the matter-enhanced spin-flavor precession
effect. We used two types of magnetic field profiles throughout the entire
Sun: Wood-Saxon shape and the Gaussian shape. We showed that for Dirac
neutrinos, the allowed regions are independent of the magnetic field
profiles for all of the magnetic moments that we used in this paper and the
allowed region in the large mixing angle (LMA) region shifted to the small
mixing angle (SMA) region as $\mu B $ value is increased. We calculated the
allowed regions at 95\% CL. We also find a limit for the electron neutrino
magnetic moment at 0.95CL so that $\mu B<0.2\times10^{-7}\mu_{B}G$ for both
magnetic field profiles at 1$\sigma$level.
\end{abstract}

\textbf{1.Introduction}\linebreak

The earlier solar neutrino experiments (Homestake and all three Gallium
experiments)$\left[ 1-4\right] $ showed deficits of the neutrino flux from
the Sun when compared to the standard solar model(SSM) predictions$\left[ 5%
\right] $.

One of the most popular solutions of this problem is MSW effect$\left[ 6,7%
\right] $. In this solution, when electron neutrinos pass through the
electronic matter, they are exposed to matter effects which can cause an
almost complete conversion of electron neutrinos to another neutrino types.

Another solution is the magnetic moment solution which we investigated in
this paper. If the neutrinos have large magnetic moments and pass through a
region with a magnetic field, the helicity of the neutrino can be flipped,
then this spin flip changes the left-handed electron neutrino to a right
handed electron neutrino that we cannot detect$\left[ 8-11\right] $.

Lim and Marciano examined$\left[ 12\right] $ the combined effect of matter
and magnetic fields on neutrino spin and flavor precession. They developed
the idea of the resonance spin flavor precession(RSFP) in 1988. In this
solution based on the RSFP mechanism, the combined effect changes the
neutrino's chirality and flavor simultaneously. Matter-enhanced spin-flavor
precession of solar neutrinos with transition magnetic moments for chlorine
and gallium experiments was investigated in detail by Baha \textit{et al }$%
\left[ 13\right] $. In recent years there have been several other studies on
RSFP investigating different aspects$\left[ 14-23\right] $. Chauhan et.al $%
\left[ 23\right] $ examined the combined action of neutrino oscillation and
spin-flavor precession (SFP). They considered the flux of electron
antineutrinos coming from the Sun as a possible observable effect of SFP,
and as a result tried to put an upper bound as $\mu B<$ $2.8\times
10^{-19}MeV$. Also to examine in detail the effect of the Sun's magnetic
field, a statistical analysis of the solar data has been performed in $[21]$
and from the minimum of chi-squares, the value of the magnetic field for
different profiles can be deduced. In addition to these, one can examine
first the allowed region of neutrino parameter space for each neutrino
experiment (Chlorine, Gallium, SK and SNO) depending on the magnetic field
strength in the Sun and secondly can find the allowed regions from combined
solar $\nu $ experiments at the same magnetic field strength, by adding the
KamLAND data to the combined solar $\nu $ data to make the analysis global.
In our present present work we followed such a path to extract a upper $\mu
B $ value.

In the present work a global analysis of the solar neutrino data combined
with the new binned KamLAND data $\left[ 24\right] $ is presented assuming
the solar neutrino deficit to be resolved by the matter-enhanced spin-flavor
precession. We calculated survival probability for two different magnetic
field profiles.

Standard least-squares analysis of solar neutrino data is often used to
obtain information on the values of the allowed regions for the oscillation
parameters namely $\Delta m^{2}$, $\tan^{2}\theta$. Large mixing angle
(LMA), small mixing angle (SMA), and low $\delta m^{2}$(LOW) regions are
usually known as MSW solutions. Besides the earlier solar neutrino
experiments Homestake $\left[ 1\right] $ and Gallium$\left[ 2-4\right] $,
SNO charged-current and neutral current $\left[ 25,26\right] $ and Super
Kamiokande $\left[ 27\right] $ results comfirmed solar neutrino deficit.

In 2002 global analysis of all solar neutrino experiments showed that the
LMA solution was the most likely solution in the neutrino parameter space$%
\left[ 28\right] $. Also data from the KamLAND reactor neutrino experiment
indicated the LMA region $\left[ 24,29\right] $. So, solar and reactor
neutrino experiments strongly pointed out the LMA region.

In our paper, we examined how the solar magnetic field in the Sun can change
this region. We found results on the allowed regions and chi-square values
for two different magnetic field profiles and for nine different values of
neutrino magnetic moments. Our results showed that this allowed region in
the LMA shifted to the SMA region for certain $\mu $B values. From these
observations we were able to put upper limits on the $\mu $B values. In
section 2, we give general information about equations that governs the
neutrino propagation assuming that neutrinos have magnetic
moments.Properties of the magnetic field profiles are given in section 3. We
give detailed statistical analysis in section 4. Finally our results and
conclusion are given in the section 5.\nolinebreak

\textbf{2. Matter-enhanced spin-flavor precession in the Sun}

For the electron neutrino, in the case of $\nu_{e_{L}}\rightarrow\nu_{e_{R}}$
precession the $\nu_{e_{L}}$ and $\nu_{e_{R}}$ interact differently with
matter. The differences in their interactions with matter suppresses
precession by splitting their degeneracy.

The evolution equation that describes the propagation throughout matter of
the two chiral components $\nu_{e_{L}}$ and $\nu_{e_{R}}$ with magnetic
moment $\mu_{e_{v}}$ is $\left[ 8,9,30\right] $

\begin{equation}
i\frac{d}{dt}\left[ 
\begin{array}{c}
\nu_{e_{L}} \\ 
\nu_{e_{R}}%
\end{array}
\right] =\left[ 
\begin{array}{cc}
V_{e}(t) & \mu B \\ 
\mu B & 0%
\end{array}
\right]\left[ 
\begin{array}{c}
\nu_{e_{L}} \\ 
\nu_{e_{R}}%
\end{array}
\right]  \label{1}
\end{equation}
where $B$ is the transverse magnetic field and $V_{e}(t)$ is the "matter"
potential that is the contribution of matter to the effective mass. In the
standard model for an unpolarized neutral medium,

\begin{equation}
V_{e}(t)=\frac{G_{f}}{\sqrt{2}}(2N_{e}-N_{n})
\end{equation}%
where $N_{e}$ and $N_{n}$ are electron and neutron number densities
respectively and the $G_{f}=1.16636\times 10^{-5}GeV^{-2}$. Also in the Sun
electron and neutron number densities are well approximated by $N_{e}\simeq
6N_{n}\simeq 2.4\times 10^{26}\exp (-r/0.09R_{\odot })/cm^{3}$. In the $%
\left[ 30\right] $, the resonance condition was found as $N_{e}\simeq
N_{n}/2 $. This neutron number density condition can not be found within the
Sun, but it can exist in a supernonova $\left[ 31,32\right] $.

Although there is no resonance region in the Sun for spin-precession, it was
shown that there is a resonance region where neutrino spin-flavor precession
may occur for a medium with changing density.

To explain the resonant spin-flavor precession, we first consider two
generations. For definiteness, we examine $\nu _{e}-\nu _{\mu }$ system. The
evolution equation for a neutrino that propagates through matter and a
transverse magnetic field $B$ is

\begin{equation}
i\frac{d}{dt}\left[ 
\begin{array}{c}
\nu_{e_{L}} \\ 
\nu_{\mu_{L}} \\ 
\nu_{e_{R}} \\ 
\nu_{\mu_{R}}%
\end{array}
\right] =\left[ 
\begin{array}{cc}
H_{L} & BM^{\dagger} \\ 
BM & H_{R}%
\end{array}
\right] \left[ 
\begin{array}{c}
\nu_{e_{L}} \\ 
\nu_{\mu_{L}} \\ 
\nu_{e_{R}} \\ 
\nu_{\mu_{R}}%
\end{array}
\right]  \label{2}
\end{equation}
where $2\times2$ submatrices are

\begin{equation}
H_{L}=\left[ 
\begin{array}{cc}
\frac{\Delta m^{2}}{2E}\sin^{2}\theta+V_{e} & \frac{\Delta m^{2}}{4E}%
\sin2\theta \\ 
\frac{\Delta m^{2}}{4E}\sin2\theta & \frac{\Delta m^{2}}{2E}%
\cos^{2}\theta+V_{\mu}%
\end{array}
\right]
\end{equation}

\begin{equation}
H_{R}=\left[ 
\begin{array}{cc}
\frac{\Delta m^{2}}{2E}\sin^{2}\theta & \frac{\Delta m^{2}}{4E}\sin2\theta
\\ 
\frac{\Delta m^{2}}{4E}\sin2\theta & \frac{\Delta m^{2}}{2E}\cos^{2}\theta%
\end{array}
\right]\text{ \ \ }M=\left[ 
\begin{array}{cc}
\mu_{ee} & \mu_{e{\mu}} \\ 
\mu_{{\mu}e} & \mu_{{\mu}{\mu}}%
\end{array}
\right]
\end{equation}

where $\theta $ is the mixing angle, $\Delta m^{2}$ is the difference of the
mass and $E$ is the neutrino energy. The matter potentials for a neutral
unpolarized medium are given as

\begin{equation}
V_{e}(t)=\frac{G_{f}}{\sqrt{2}}(2N_{e}-N_{n})\text{ \ \ \ \ \ \ }V_{\mu }=-%
\frac{G_{f}}{\sqrt{2}}N_{n}
\end{equation}

\textbf{3.Magnetic Field Profiles\nolinebreak}

\begin{figure}[ht]
\centering \includegraphics[width=4in]{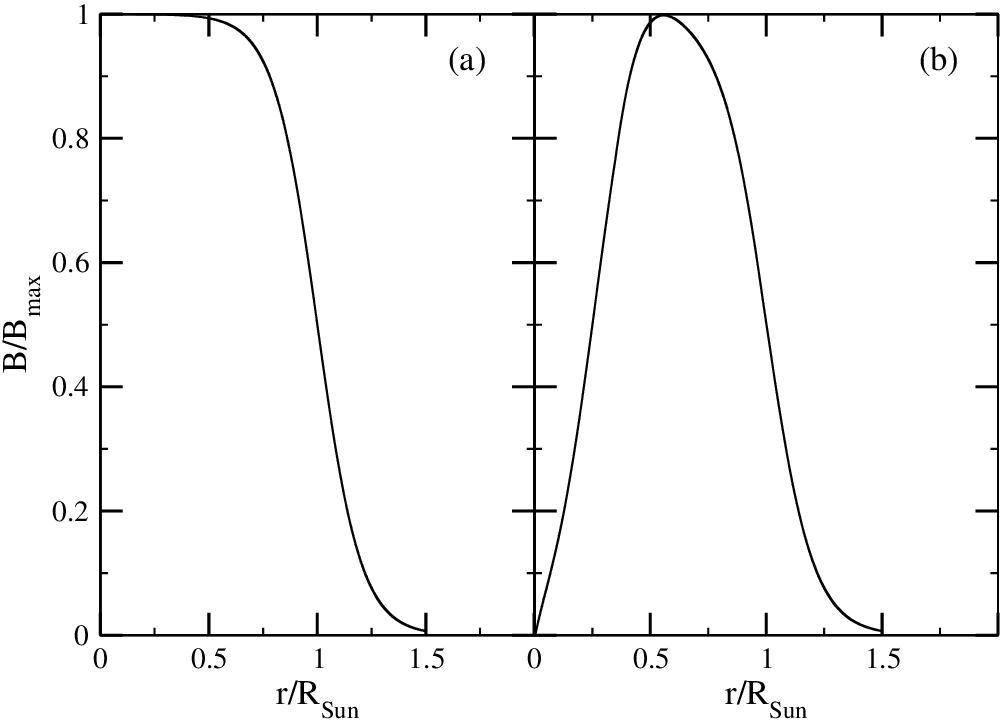}
\caption{Magnetic field profiles: (a) Wood-Saxon shape; (b) Gaussian shape}
\label{fig:figure1.eps}
\end{figure}

In our analysis, we used two types of magnetic profiles.

First, we took the magnetic field profile to be a Wood-Saxon shape of the
form

\begin{equation}
B(r)=\frac{B_{0}}{1+\exp[10(r-R_{\odot})/R_{\odot}]}  \label{3}
\end{equation}
where $B_{0}$ is the strength of the magnetic field at the center of the Sun.

The second magnetic field profile that we used is a Gaussian shape. Altough
there are many other profiles $\left[ 14,17-19,21,23\right] $ we considered
only those two as typical for our purpose as shown figure 1. The effects of
these different profiles have been discussed in the conclusion section.

\textbf{4. Statistical Analysis}

In the literature, there is a common way often called $\chi^{2}$ analysis to
find the values of the neutrino oscillation parameters $\Delta m^{2}$, $%
\tan^{2}\theta$ and to calculate the confidence levels of allowed regions
and the goodness of a fit $\left[ 33-36\right] $. In our analysis, we use
"covariance approach" to find the allowed regions mentioned above. By this
method, one minimizes the least-squares function

\begin{equation}
\chi _{_{\odot }}^{2}=\sum_{i_{1},i_{2}}^{N_{\exp }}(R_{i_{1}}^{(\exp
)}-R_{i_{1}}^{(thr)})(V^{-1})_{i_{1}i_{2}}(R_{i_{2}}^{(\exp
)}-R_{i_{2}}^{(thr)})  \label{4}
\end{equation}%
where $V^{-1}$ is the inverse of the covariance matrix of experimental and
theoretical uncertainties, $R_{i}^{(\exp )}$ is event rate calculated in the 
$i$th experiment and $R_{i}^{(thr)}$ is the theoretical event rate for $i$th
experiment. The indices indicate the solar neutrino experiments: $%
i,i_{1},i_{2}=1,...,N_{\exp }$ with $N_{\exp }=4$.

\begin{equation}
V_{i_{1}i_{2}}=V_{i_{1}i_{2}}(\exp)+V_{i_{1}i_{2}}(thr)  \label{5}
\end{equation}

\begin{equation}
V_{i_{1}i_{2}}(\exp )=\delta _{i_{1}i_{2}}\sigma _{i_{1}}^{\exp }\sigma
_{i_{2}}^{\exp }  \label{6}
\end{equation}%
where $\sigma _{i}^{\exp }$ are the experimental uncertainties for $i$th
experiment.

\begin{align}
V_{i_{1}i_{2}}(thr) &
=\delta_{i_{1}i_{2}}\sum_{j_{1}=1}^{8}R_{j_{1}i_{1}}^{(thr)^{2}}\Delta\ln
C_{j_{1}i_{1}}^{(thr)^{2}}+  \label{7} \\
&
\sum_{j_{1},j_{2}=1}^{8}R_{j_{1}i_{1}}^{(thr)^{2}}R_{j_{2}i_{2}}^{(thr)^{2}}%
\sum_{k=1}^{12}\alpha_{j_{1}k}\alpha_{j_{2}k}(\Delta\ln X_{k})^{2}  \notag
\end{align}
where the indices $j_{1}$,$j_{2}=1,...,8$ indicate solar neutrino fluxes
produced in the eight thermonuclear reactions in the Sun: pp, pep, hep, $%
^{7} $Be, $^{8}$B, $^{13}$N, $^{15}$O, $^{17}$F, respectively. The index
k=1,...,12 denotes the input astrophysical parameters $X_{k}$ in the
Standard Solar Model(SSM), on which SSM neutrino fluxes $\Phi_{j}^{SSM}$
depend. The logarithmic derivatives

\bigskip%
\begin{equation}
\alpha_{jk}=\frac{\partial\ln\Phi_{j}^{SSM}}{\partial\ln X_{k}}
\end{equation}
govern the uncertainties of the neutrino fluxes $\Phi_{j}^{SSM}$. $\Delta\ln
X_{k}$ and $\Delta\ln C_{ji}^{(thr)}$ are 1$\sigma$ relative uncertainties
of SSM input parameters and the energy-averaged cross section( $%
C_{ji}^{(thr)}$), respectively.

Theoretical event rates for the radio-chemical experiments, chlorine
experiments(Homestake) and gallium experiments(SAGE,GALLEX,GNO), can be
found as

\begin{equation}
R_{ij}^{(thr)}=\int dE\phi_{i}\left( E\right) \sigma_{j}(E)P_{ij}(\nu
_{e}\rightarrow\nu_{e},E)
\end{equation}
such that

\begin{equation}
R_{j}^{(thr)}=\sum_{i=1}^{8}R_{ij}^{(thr)}
\end{equation}%
where $\phi _{i}\left( E\right) $ is the flux at energy E coming from $i$th
reaction and $\sigma _{j}(E)$ is the cross section for detector j.

Solar neutrinos are observed in SK via the Cerenkov light from the
neutrino-electron scattering (ES) reaction:

\begin{equation*}
\nu_{x}+e^{-}\rightarrow\nu_{x}+e^{-}
\end{equation*}
where $\nu_{x}$ can be $\nu_{e}$ or $\nu_{\mu}$.

Since SNO is heavy water-Cerenkov detector, it observes solar neutrinos by
charged-current(CC) and neutral-current(NC) in addition to ES

\begin{align*}
\nu_{e}+d & \rightarrow p+p+e^{-}\text{ \ \ \ \ \ \ \ \ \ \ \ \ \ }(CC) \\
\nu_{x}+d & \rightarrow n+p+\nu_{x}\text{ \ \ \ \ \ \ \ \ \ \ \ \ \ }(NC)
\end{align*}
The \ Cerenkov light is generated by recoiling electron from the ES and CC
reactions.

Because of higher threshold energy of SK and SNO experiments($%
T_{threshold}>5 $ $MeV$), they are sensitive to only $^{8}$B and hep
neutrinos. Due to the small flux of hep neutrinos we completely neglect its
contribution to the total rates.

For SK and SNO, the theoretical event rates from ES:

\begin{align}
R_{SK}^{ES} & =\int dE\phi_{i}\left( E\right)
\{\sigma_{\nu_{e}}^{_{j}}(E)P_{j}(\nu_{e}\rightarrow\nu_{e},E,t)+ \\
& \text{ \ \ \ \ \ \ \ \ \ \ \ \ \ \ \ \ \ \ }\sigma_{\nu_{x}}^{_{j}}(E)%
\left[ 1-P_{j}(\nu_{e}\rightarrow\nu_{e},E,t)\right] \}  \notag
\end{align}
where

\begin{equation}
\sigma _{\nu _{e}(\nu _{x})}^{_{j}}(E)=\int_{T_{\min }}^{T_{\max }}\frac{%
d^{2}\sigma }{dTdE}dT
\end{equation}%
Here $T$ is the kinetic energy of the recoiling electron. $T_{\min
}=E_{threshold}-m_{e}$ with $E_{threshold}=5.5$ MeV and $T_{\max
}=2E/(2E+m_{e})$ are the minimum and maximum kinetic energy of the recoiling
electron, respectively. The differantial cross sections for $\nu _{e}-e^{-}$
and $\nu _{x}-e^{-}$ scatterings are $[37]$

\begin{equation}
\frac{d^{2}\sigma}{dTdE}%
=\sigma_{e}[g_{L}^{2}+g_{R}^{2}(1-T/E)^{2}-g_{L}g_{R}(T/E^{2})]
\end{equation}
with

\begin{equation*}
g_{L}=(\pm\frac{1}{2}+\sin^{2}\theta_{W})\text{ \ \ \ \ \ \ }g_{R}=\sin
^{2}\theta_{W}\text{\ \ \ \ \ \ \ \ }
\end{equation*}
The upper sign is for $\nu_{e}-e^{-}$ scattering and the lower sign for $%
\nu_{x}-e^{-}$ scattering; x can be $\mu$ or $\tau.$ The cross section
factor is

\begin{equation*}
\sigma_{e}=\frac{2G_{F}^{2}m_{e}^{2}}{\pi h^{4}}=88.083\times10^{-46}cm^{2}
\end{equation*}

For SNO, in addition to the ES, the theoretical event rates come from the CC
and NC reactions

\begin{align}
R^{CC} & =\int dE\phi_{i}\left( E\right) \sigma_{CC}^{_{j}}(E)P_{j}(\nu
_{e}\rightarrow\nu_{e},E,t) \\
R^{NC} & =\int dE\phi_{i}\left( E\right) \sigma_{NC}^{_{j}}(E)
\end{align}
Total event rates are

\begin{equation*}
R^{SNO}=R^{CC}+R^{NC}
\end{equation*}%
Because of its substantially lower cross section compered to the other two
reactions', we don't take into account $R^{ES}$.

\begin{figure}[th]
\centering \includegraphics[width=4in]{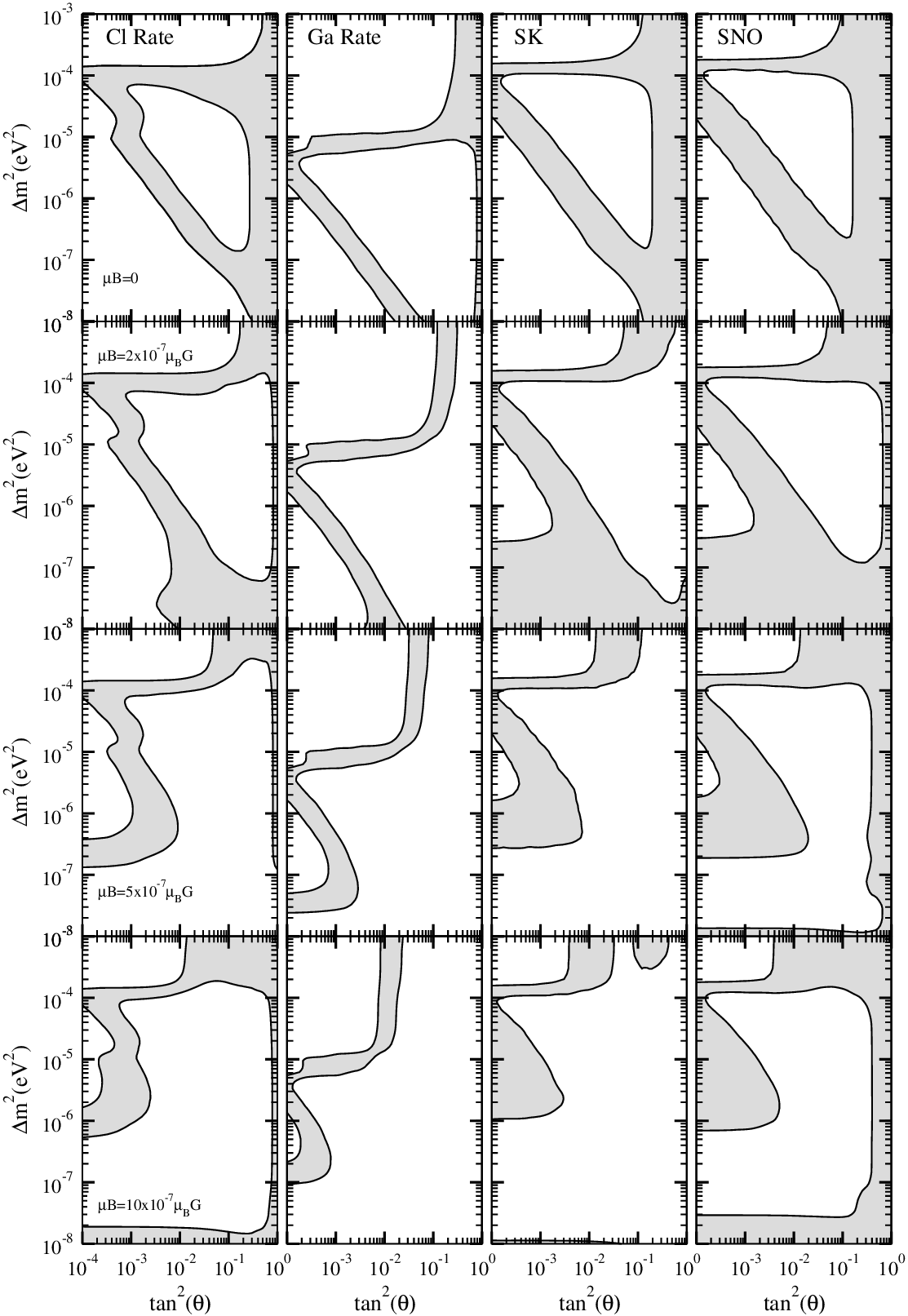}
\caption{The allowed regions of neutrino parameter space for each solar
neutrino experiment seperately at $\protect\mu B=0,2,5,10\times 10^{-7}%
\protect\mu _{B}G$ and at 95\% CL. Each column and row are for the same
experiment and at the same $\protect\mu B$ value,respectively(e.g. in the
second row at third column, an allowed region for SK experiment at $\protect%
\mu B=2\times 10^{-7}\protect\mu _{B}G$ is seen)}
\label{fig:figure2.eps}
\end{figure}

We took fluxes and cross sections for event rates and error matrixes from
Bahcall $[37$, homepage$]$.

\begin{figure}[ht]
\centering \includegraphics[width=4in]{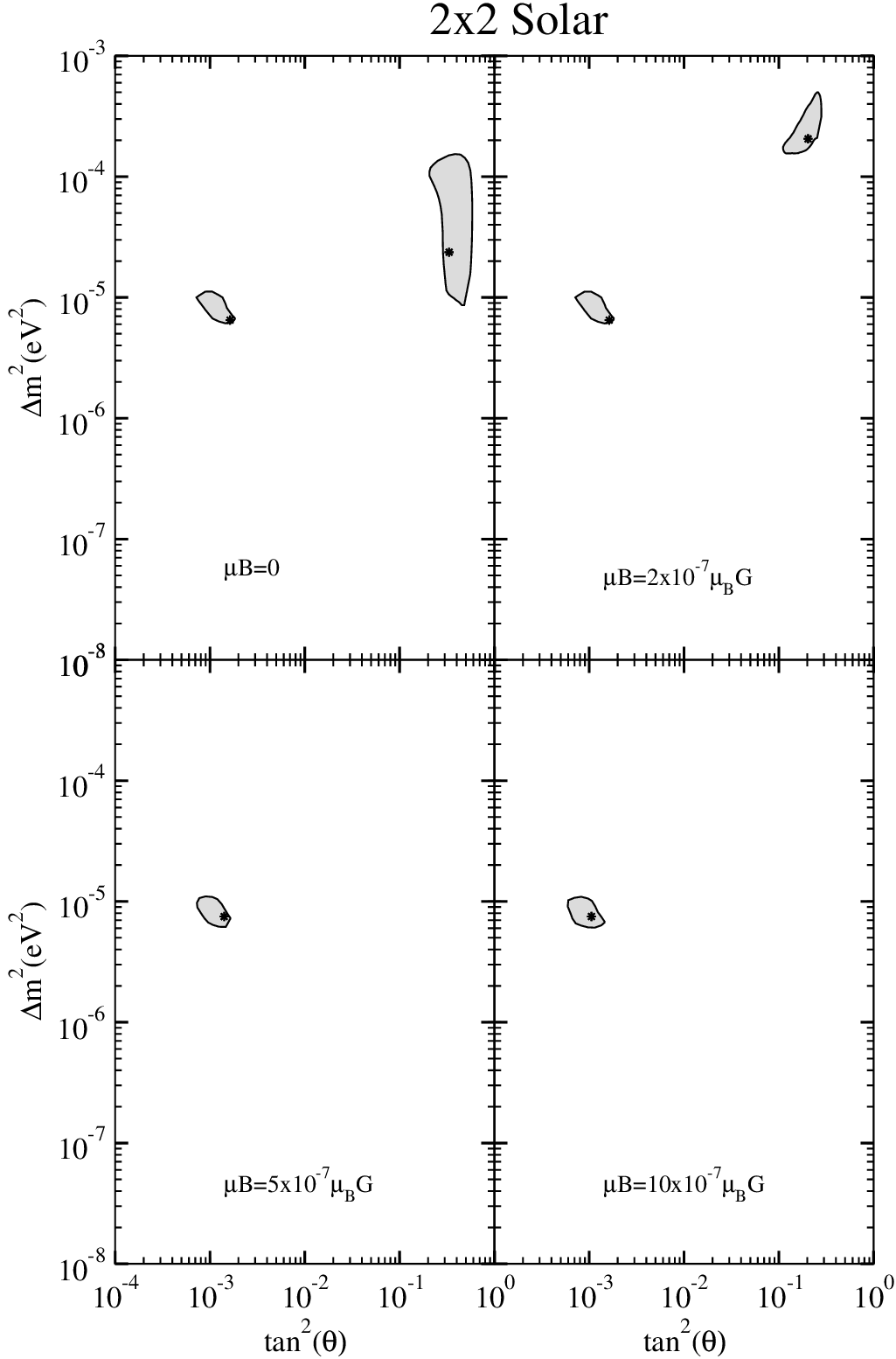}
\caption{The allowed regions from combined solar neutrino experiments(
chlorine, all three gallium, SK\ and SNO experiments) at the same $\protect%
\mu B$ values and CL as in figure 2. Stars indicate the local best-fit
points.}
\label{fig:figure3.eps}
\end{figure}

We need to calculate $\chi_{KamLAND}^{2}$ for the global analysis

\begin{equation}
\chi_{gl}^{2}=\chi_{_{\odot}}^{2}+\chi_{_{KamLAND}}^{2}
\end{equation}
For this purpose we note that, the goal of the KamLAND is to search for
antineutrinos emitted from distant power reactor through the reaction

\begin{equation}
p+\overline{\nu }_{e}\rightarrow n+e^{+}
\end{equation}%
As in $[38]$, the antineutrinos' energy spectrum is given by

\begin{equation}
\frac{dN_{\overline{\nu }_{e}}^{j}}{dE}\propto \exp (a_{0}+a_{1}E+a_{2}E^{2})
\end{equation}%
where $j=1,2,3,4$ corresponding to the four isotopes $^{235}$U, $^{239}$P, $%
^{238}$U, $^{241}$Pu and the fitted values of the $a_{k}$ are given in
detail in $[39]$ for each isotope.

In the case of two generations, the survival probability for the electron
antineutrinos coming from the j$^{th}$ reactor

\begin{equation}
P(\overline{\nu }_{e}\rightarrow \overline{\nu }_{e})=1-\sin ^{2}2\theta
\sin ^{2}(\frac{1.27\Delta m^{2}(eV^{2})d_{j}(km)}{E(GeV)})
\end{equation}%
here $d_{j}$ is the reactor-detector distance. The number of expected events
for each bin at KamLAND is

\begin{align}
N_{i}^{thr}(t,E,\Delta m^{2},\sin ^{2}2\theta )& =\eta N_{p}{\displaystyle%
\int }dE_{V}{\displaystyle\int }dE_{e}R(E_{V},E_{e})\times \\
& \text{ \ \ \ \ \ \ }\sum_{j}\frac{S_{j}}{4\pi d_{j}^{2}}\sigma (E_{\nu })P(%
\overline{\nu }_{e}\rightarrow \overline{\nu }_{e},E_{\nu })  \notag
\end{align}%
here $N_{p}$ is the number of free protons in the fiducial volume of
detector and $\eta $ is the efficiency which are given in $[24]$. $S_{j}$ is
the initial energy spectrum of reactor $j$ calculated using the thermal
power and the isotropic composition of each detector that is given in
detailed in $[39]$ and $\sigma (E_{\nu })$ is the lowest cross section

\begin{equation}
\sigma(E_{\nu})=\frac{2\pi^{2}}{m_{e}^{5}f\tau_{n}}p_{e}E_{e}
\end{equation}
where $f=1.69$ is the integrated Fermi function for neutron, $m_{e}$ is the
positron mass, $\tau_{n}$ is the neutron lifetime, $p_{e}$ and $E_{e}$ are
the positron momentum and energy, respectively. So that we have

\begin{equation}
E_{e}=E_{\nu}-1.293\text{ MeV}
\end{equation}

The energy resolution function $R(E_{V},E_{e})$ which depends on visible
energy $\left( E_{V}\right) $ and true positron energy $\left( E_{e}\right) $
is given by

\begin{equation}
R(E_{V},E_{e})=\frac{1}{\sqrt{2\pi\sigma_{0}^{2}}}e^{-\frac{\left(
E_{V}-E_{e}+m_{e}\right) ^{2}}{2\sigma_{0}^{2}}}
\end{equation}
in which $E_{V}=E_{e}+m_{e}$ and $\sigma_{0}=6.2\%\sqrt{E_{e}}$ $[24]$.

The new KamLAND data results were reported at $[24]$ in 13 bins above the
threshold which is 2.6 MeV. To find $\chi^{2}$ for KamLAND spectrum data,
due to the low statistics, we use $\chi^{2}$ assuming a Poisson distribution
given by

\begin{equation}
\chi _{KL-Spect}^{2}=\underset{i}{\dsum }\left[ 2\left( \kappa
N_{i}^{thr}-N_{i}^{\exp }\right) +2N_{i}^{\exp }\ln (\frac{N_{i}^{\exp }}{%
\kappa N_{i}^{thr}})\right] +\frac{(\kappa -1)^{2}}{\sigma _{sys}^{2}}
\end{equation}%
where sum is over the KamLAND spectral bins, $\sigma _{sys}^{2}$ is the
systematic uncertainty taken to be $6.5\%$ and $\kappa $ is allowed to vary
freely.

\begin{figure}[th]
\begin{center}
\includegraphics[width=4in]{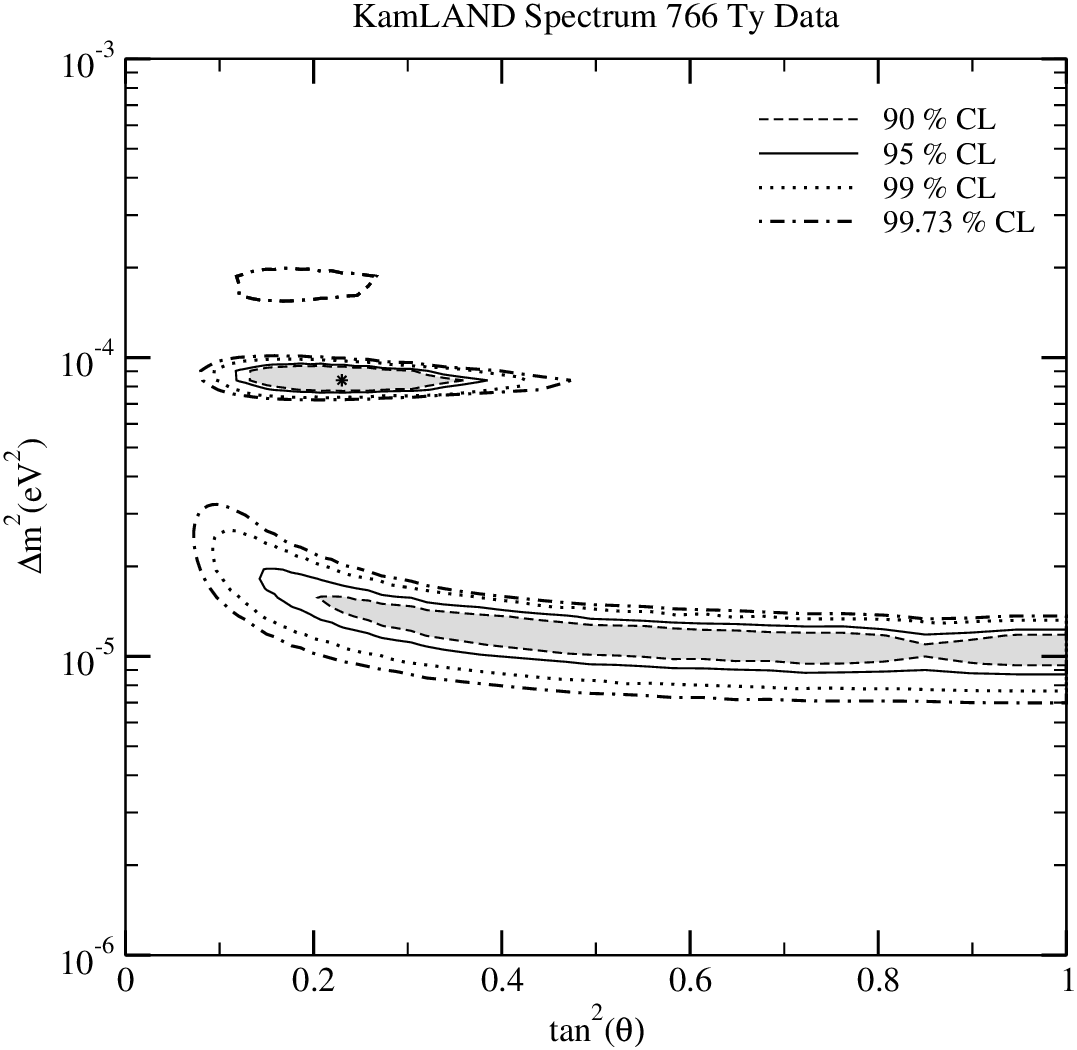}
\end{center}
\caption{Allowed regions from the new KamLAND spectrum at different
confidence levels. The star indicates the best fit point.}
\label{fig:figure4.eps}
\end{figure}

\textbf{5.Results and conclusions}

In our calculations, we assumed that the magnetic field extends over the
entire Sun for either Wood-Saxon shape or Gaussian shape. We use the
neutrino spectra from the Standard Solar Model of Bahcall and his
collaborators $\left[ 40\right] $. In our analysis we calculated allowed
regions at 95\% confidence level.

\begin{figure}[th]
\centering \includegraphics[width=4in]{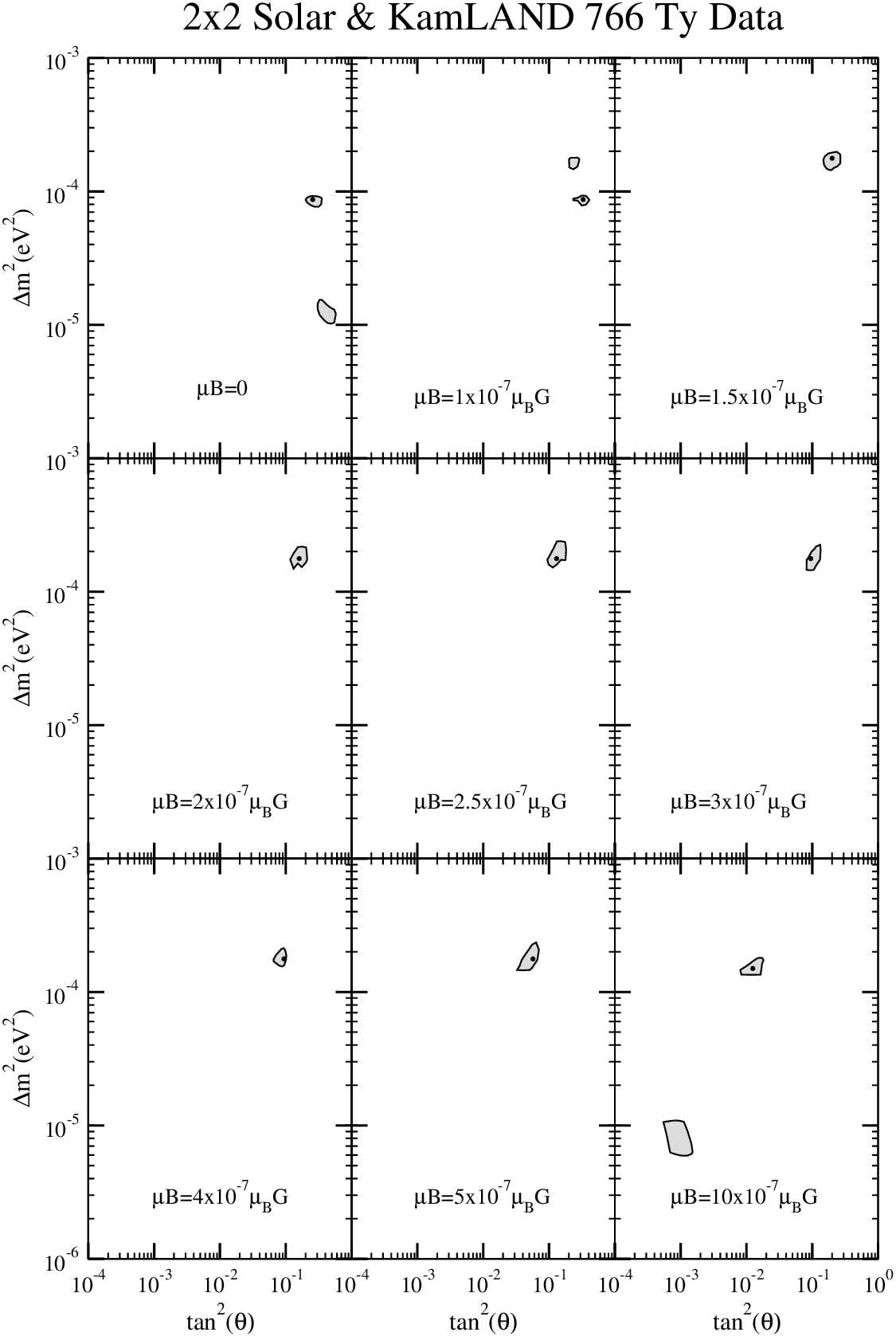}
\caption{Allowed regions from the combined solar+new KamLAND spectral
analysis at different $\protect\mu B$ values and at 95\% CL. Stars indicate
the best fit points.}
\label{fig:figure5.eps}
\end{figure}

We first considered only solar neutrino data. In our statistical analysis,
we used the covariance approach. We show the allowed regions of neutrino
parameter space in figure 2 for each solar neutrino experiment seperately at
four $\mu B$ values using Wood-Saxon field profiles. In figure 2 each column
and row are for the same experiment and at the same $\mu B$ value
respectively(e.g. in the second row at the third column, an allowed region
for SK experiment at $\mu B=2\times 10^{-7}\mu _{B}G$ is seen).

In figure 3, we displayed the allowed regions from combined solar neutrino
experiments at the same values of $\mu B$ in figure 2.

After we investigated the combined solar data at different $\mu B$ values,
we looked for allowed regions at different confidence levels from new binned
KamLAND data $[24]$ in figure 4. We show the allowed regions from our global
analysis combining solar and new KamLAND data in figure 5 for nine different 
$\mu B$ values($0,1,1.5,2,2.5,3,4,5,10\times 10^{-7}\mu _{B}G$ ). In that
figure, our results showed that the allowed regions in the LMA shifted to
the SMA region as $\mu B$ value is increased. This shift enables us to put
an upper limit on the value of $\mu B$ , since the latest experimental data
prominently indicate the LMA region (namely excluding SMA and the others).

\bigskip

\textbf{Table1}. Best fit points of global analysis for both magnetic field
profiles.

\begin{tabular}{ccccc}
\hline
$\mu B$ $(${\small \ }$\times10^{-7}${\small \ }$\mu_{B}G)$ & $\Delta
m^{2}(eV^{2})$ & tan$^{2}\theta$ & ($\chi_{\min}^{2}$)$^{Wood-Saxon}$ & ($%
\chi_{\min}^{2}$)$^{Gaussian}$ \\ \hline
0 & 8.7$\times10^{-5}$ & 0.26 & 26.38 & 26.38 \\ 
1 & 8.7$\times10^{-5}$ & 0.33 & 35.4 & 35.64 \\ 
1.5 & 1.77$\times10^{-4}$ & 0.20 & 38.6 & 38.8 \\ 
2 & 1.77$\times10^{-4}$ & 0.16 & 41.13 & 41.16 \\ 
2.5 & 1.77$\times10^{-4}$ & 0.13 & 43.81 & 44.02 \\ 
3 & 1.77$\times10^{-4}$ & 0.095 & 42.47 & 42.92 \\ 
4 & 1.77$\times10^{-4}$ & 0.095 & 42.59 & 42.45 \\ 
5 & 1.77$\times10^{-4}$ & 0.057 & 49.39 & 49.73 \\ 
7 & 1.53$\times10^{-4}$ & 0.022 & 53.78 & 54.13 \\ 
10 & 1.53$\times10^{-4}$ & 0.012 & 55.5 & 55.85 \\ \hline
\end{tabular}

\bigskip

We next calculated the global analysis again for a magnetic profile of
Gaussian-shape to show how the magnetic profile in the sun effects the
allowed regions and the minimum chi-squares. The results of our global
analysis for both magnetic field profiles are given in table 1. As can be
seen from the table there are no appreciable differences between the effects
of the two magnetic field profiles on the allowed regions and chi-squares
values.

\begin{figure}[th]
\centering \includegraphics[width=3in]{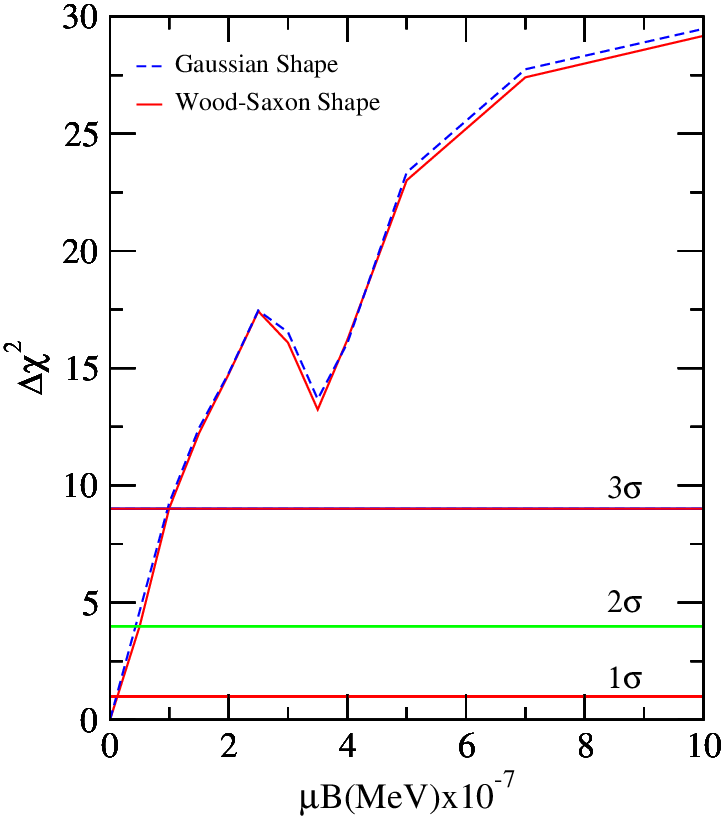}
\caption{Projection of the global $\Delta \protect\chi ^{2}$ function on $%
\protect\mu B$. The solid line and short-dashed lines represent the
calculation using Wood-Saxon shape of magnetic profile and Gaussian shape of
magnetic profile, respectively.}
\label{fig:figure6.eps}
\end{figure}

Finally in figure 6 we present projection of the global $\Delta \chi ^{2}$
function on the $\mu B$ \ to place a limit on $\mu B$ for the two magnetic
field profiles. From this figure we found a general limit, since the graphes
for the two magnetic profiles almost coincide: $\mu B<$ $0.2\times
10^{-7}\mu _{B}G$, $0.5\times 10^{-7}\mu _{B}G$, $1.0\times 10^{-7}\mu _{B}G$
for $1\sigma ,2\sigma ,3\sigma $ limit, respectively. On the other hand
there is intensive work related to the electromagnetic properties of the
neutrinos. Since neutrinos are known to be massive they can get tiny
magnetic moments even in the SM also in its extensions. Astrophysical
considerations $[41]$ put strong constraints on the magnetic moment of the
neutrinos as $\mu _{\nu }<10^{-12}\mu _{B}$, however these bounds are model
dependent. Also the reactor experiments bring less restrictive limits of $%
\mu _{\nu }<1.3\times 10^{-10}\mu _{B}$ $[42,43]$. An analysis of the
Super-Kamiokande solar neutrino data $[44]$ enables the authors to put again
a similar limit of $\mu _{\nu }<1.5\times 10^{-10}\mu _{B}$. As for the
magnetic field of the Sun, from the existing observations an upper limit of $%
10^{7}G$ at the core and a maximum magnitude of $3\times 10^{5}G$ at the
bottom of the convective zone are usually taken in the literature [21, 22].
Hence the above restriction according to our analysis on the $\mu B$ value
agrees with the present limits of $\mu _{\nu }$ and $B_{Sun}$.

Also the consideration of the two magnetic field profiles seem to be
satisfactory at this level, because there are no differences between their
effects on the allowed regions and the chi-square values for both of them.
Actually the whole analysis can be repeated for many other field profiles $%
[14,17-19,21,23]$ in order to better investigate the detailed behaviour of
the magnetic field in the Sun. This could be a topic of a future work.

\bigskip

\bigskip

\textbf{Acknowledgments}

We would like to thank to Prof.Dr. A. Baha Balantekin for his suggestion of
this research topic and for his kind help. We are gratefull to the referees
for their clarifying suggestions. One of the authors, D. Yilmaz, also thanks
TUBITAK (The Scientific and Technical Research Council of Turkey) for the
grant given to support his research at University of Wisconsin-Madison.

\bigskip

\textbf{References}

\bigskip

$\left[ 1\right] $ \ \ Cleveland B T \textit{et al }1998 \textit{Astrophys. J%
}. \textbf{496} 505

$\left[ 2\right] $ \ \ Abdurashitov J N \textit{et al} (SAGE Collaboration)
2002 \textit{J. Exp. Theor. Phys.}\textbf{\ 95} 181

\ \ \ \ \ \ \ Abdurashitov J N \textit{et al} 2002 \textit{Zh. Eksp. Teor.
Fiz. }\textbf{122} 211 \textit{(Preprint astro-ph/0204245)}

$\left[ 3\right] $ \ \ Hampel W \textit{et al }(GALLEX Collaboration) 1999 
\textit{Phys. Lett. B} \textbf{447} 127

$\left[ 4\right] $ \ \ Altmann M \textit{et al} (GNO Collaboration) 2000 
\textit{Phys. Lett. }B \textbf{490} 16 \textit{(Preprint hep-ex/0006034)}

$\left[ 5\right] $ Bahcall J N, Huebner W F, Lubow S H, Parker P D , and
Ulrich R K , \textit{Rev. Mod. Phys.} \textbf{54}, 767 (1982)

$\left[ 6\right] $ Wolfenstein L, \textit{Phys. Rev. D} \textbf{17}, 2369
(1978) ; \textbf{20}, 2634 (1979).

$\left[ 7\right] $ Mikheyev S P and Smirnov A Yu , \textit{Nuovo Cimento} C 
\textbf{9}, 17 (1986) ; \textit{Yad. Fiz.} \textbf{42}, 1441 (1985) ; [Sov. 
\textit{J. Nucl. Phys.} \textbf{42}, 913 (1985)].

$\left[ 8\right] $ Okun L B, Voloshin M B, and Vysotsky M I,\textit{\ Yad.
Fiz. }\textbf{44}, 677 (1986) ; [\textit{Sov. J. Nucl. Phys}. \textbf{44},
440 (1986)].

$\left[ 9\right] $ Akhmedov E Kh , \textit{Phys. Lett.} B \textbf{213}, 64
(1988); E. Kh. Akhmedov and M. Yu. Khlopov, \textit{Mod. Phys. Lett.} A 
\textbf{3}, 451 (1988).

$\left[ 10\right] $ Barbieri R and Fiorentini G, \textit{Nucl. Phys.} B 
\textbf{304}, 909 (1988).

$\left[ 11\right] $ Bethe H A, \textit{Phys. Rev. Lett.} \textbf{63}, 837
(1989) .

$\left[ 12\right] $ Lim C S and Marciano W J, \textit{Phys. Rev.} D \textbf{%
37}, 1368 (1988)

$\left[ 13\right] $ Balantekin A B, Hatchell P J, Loreti F, \textit{Phys.
Rev.} D \textbf{41} 3583 (1990)

$\left[ 14\right] $ Pulido J, 2002 \textit{A. High Energy Phys. }AHEP2003/046

$\left[ 15\right] $ Chauhan C B, Pulido J, 2004 \textit{Preprint}
hep-ph/0402194

$\left[ 16\right] $ Lim C S, 2001 \textit{Kashiwa 2001, Neutrino Oscillation
and Their Origin} 64

$\left[ 17\right] $ Chauhan C B, Pulido J, 2002 \textit{Preprint}
hep-ph/0206193

$\left[ 18\right] $ Chauhan C B, 2002 \textit{Preprint} hep-ph/0204160

$\left[ 19\right] $ Bykov A A, Popov V Y, Rashba T I, Semikoz V B, 1999 
\textit{Preprint} hep-ph/0002174

$\left[ 20\right] $ Derkaoui J, Tayalati Y, 1999 \textit{Preprint}
hep-ph/9909512

$\left[ 21\right] $ Akhmedov E Kh, Pulido J, 1999 \textit{Preprint}
hep-ph/9907399

$\left[ 22\right] $ Akhmedov E Kh, Pulido J, 2002 \textit{Preprint}
hep-ph/0209192

$\left[ 23\right] $ Chauhan C B, Pulido J, Torrente-Lujan E, 2003 \textit{%
Preprint} hep-ph/0304297

$\left[ 24 \right] $ \ \ Araki T \textit{et al} (KamLAND Collaboration) 2004 
\textit{Preprint} hep-ph/0406035

$\left[ 25\right] $ \ \ Ahmad Q R \textit{et al} (SNO Collaboration) 2001 
\textit{Phys. Rev. Lett}. \textbf{87 }071301 \textit{(Preprint
nucl-ex/0106015)}

$\left[ 26\right] $ \ \ Ahmad Q R \textit{et al }(SNO Collaboration) 2002 
\textit{Phys. Rev. Lett}. \textbf{89} 011301 \textit{(Preprint
nucl-ex/0204008) }

$\left[ 27\right] $ \ \ Fukuda S \textit{et al }(Super-Kamiokande
Collaboration) 2001 \textit{Phys. Rev. Lett.} \textbf{86} 5651 \textit{%
(Preprint hep-ex/0103032) }

\ \ \ \ \ \ \ Fukuda S \textit{et al }(Super-Kamiokande Collaboration) 2001 
\textit{Phys. Rev. Lett}.\textbf{\ 86} 5656 \textit{(Preprint hep-ex/0103033)%
}

$\left[ 28\right] $ \ \ Bahcall J N, Gonzalez-Garcia M C and Pena-Garay C
2002 \textit{J. High Energy Phys. }JHEP 07(2002)054 \textit{(Preprint
hep-ph/0204314) }

$\left[ 29\right] $ Eguchi K \textit{et al }(KamLAND Collaboration) 2002 
\textit{Preprint} hep-ex/0212021.

$\left[ 30\right] $ Balantekin A B, Fricke S H , and Hatchell P J, \textit{%
Phys. Rev}. D \textbf{38}, 935 (1988) .

$\left[ 31\right] $ Voloshin M B , Pis'ma \textit{Zh. Eksp. Teor. Fiz.}%
\textbf{\ 47}, 421 (1988) ; [JETP \textit{Lett}. \textbf{47}, 501 (1988)]; 
\textit{Phys. Lett}. B \textbf{209}, 360 (1988).

$\left[ 32\right] $ M. Leurer and J. Liu, \textit{Phys. Lett.} B \textbf{219}%
, 304 (1989).Giuliani and S. Ranfone, \textit{Nucl. Phys}. B\textbf{325},
724 (1989).

$\left[ 33\right] $ Feldman G J and Cousins R D 1998 \textit{Phys. Rev.} D 
\textbf{57 }3873 \textit{(Preprint physics/9711021)}

$\left[ 34\right] $\ Fogli G L, Lisi E, Marrone A, Montanino D and Palazzo A
2002 \textit{Phys. Rev.} D\textbf{\ 66} 053010 \textit{(Preprint
hep-ph/0206162) }

$\left[ 35\right] $ Garzelli M V and Giunti C 2002 \textit{Astropart. Phys}. 
\textbf{17} 205 \textit{(Preprint hep-ph/0007155)}

\ \ \ \ \ \ \ Garzelli M V and Giunti C 2002 \textit{Phys. Rev. }D \textbf{65%
} 093005 \textit{(Preprint hep-ph/0111254)}

\ \ \ \ \ \ \ Garzelli M V and Giunti C 2001\textit{\ J. High Energy Phys. }%
JHEP12(2001)017 \textit{(Preprint hep-ph/0108191) }

$\left[ 36\right] $\ Gonzalez-Garcia M C and Nir Y 2002 \textit{Preprint }%
hep-ph/0202058

$\left[ 37\right] $ Bahcall J N 1989\textit{\ Neutrino Astrophysics}
(Cambridge: Cambridge University Press)

$\left[ 38\right] $ Murayama H and Pierce A, \textit{Phys. Rev.} D \textbf{65%
} 013012 \textit{(Preprint hep-ph/0012075) }

$\left[ 39\right] $ Bandyopadhyay A, Choubey S, Goswami S, Gandhi R, Roy D
P, \textit{J. Phys.} G \textbf{29}, 665 (2003)

$\left[ 40\right] $ \ \ Bahcall J N, Pinsonneault M H and Basu S 2001 
\textit{Astrophys. J.} \textbf{555} 990 \textit{(Preprint astro-ph/0010346) }

$\left[ 41\right] $ \ \ Raffelt G G 1990 \textit{Phys. Rev. Lett. } \textbf{%
64} 2856; 1999 \textit{Phys. Rep. } \textbf{320} 319; Castellani V,
Degl'Innocenti S 1993 \textit{Astrophys. J.} \textbf{402} 574

$\left[ 42\right] $ \ \ MUNU Collaboration 2002 \textit{Preprint}
hep-ex/0304011

$\left[ 43\right] $ \ \ Li H B \textit{et al} 2003 \textit{Phys. Rev. Lett. }
\textbf{90} 131802

$\left[ 44\right] $ \ \ Beacom J F, Vogel P 1999 \textit{Phys. Rev. Lett. } 
\textbf{83} 5222

\end{document}